\documentclass{optica-article}

\journal{opticajournal} 

\articletype{Research Article}

\usepackage{lineno}
\usepackage[acronym]{glossaries}
\usepackage{float}
\usepackage{breqn}
\usepackage{comment}
\usepackage{svg}
\usepackage{amsmath}
\usepackage{multirow}

\newcommand{\e}[1]{\times 10^{#1}}

\newacronym{smf}{SMF}{single-mode fibre}
\newacronym{fwhm}{FWHM}{full width at half maximum}
\newacronym{fso}{FSO}{free-space optical}
\newacronym{rms}{RMS}{root mean square}
\newacronym{tt}{TT}{tip/tilt}
\newacronym{psd}{PSD}{power spectral density}

\begin{document}

\title{Optimal design of small aperture optical terminals for free-space links}

\author{Alex M. Frost,\authormark{1,*} Benjamin P. Dix-Matthews,\authormark{1}, Shane M. Walsh \authormark{1}, David R. Gozzard \authormark{1}, and Sascha W. Schediwy \authormark{1}}

\address{\authormark{1}International Centre for Radio Astronomy Research, The University of Western Australia, Perth 6009, Western Australia, Australia}

\email{\authormark{*}alex.frost@research.uwa.edu.au} 

\begin{abstract*}We present the generalised design of low-complexity, small aperture optical terminals intended for kilometre-scale, terrestrial, free-space laser links between fixed and dynamic targets. The design features single-mode fibre coupling of the free-space beam, assisted by a fast-steering, tip/tilt mirror that enables first-order turbulence suppression and fine target tracking. The total power throughput over the free-space link and the scintillation index in fibre are optimised. The optimal tip/tilt correction bandwidth and range, aperture size, and focal length for a given link are derived using analytical atmospheric turbulence modelling and numerical simulations.

\end{abstract*}

\section{Introduction}

The use of optical frequencies for free-space signal transfer has become desirable in the pursuit of wireless high-speed communications and stable frequency transfer, where the bandwidth and stability of current radio- and microwave-frequency techniques are now major bottlenecks \cite{Hauschild2012, Kaushal2017}. \Gls{fso} signal transfer generally consists of transmitter and receiver terminals separated by some amount of atmosphere, through which a laser beam propagates. The terminals will often be in relative motion, such as a link between a fixed ground-station and an aircraft or sea vessel. Furthermore, atmospheric beam propagation is hindered by turbulence, which imparts deflections on the beam, and perturbs its wavefront. These effects scale with frequency \cite{Strohbehn1978} and are detrimental to the applications of \gls{fso} links, stemming from the large power losses and fluctuations they cause. If the beam must be coupled into \gls{smf} -- a requirement for coherent data communications and metrology applications \cite{Ip2008, McSorley2024} -- these power fluctuations are exacerbated. Hence, the design of optical terminals for operation over \gls{fso} links involves analysis of the expected turbulence conditions, the power losses they will incur, and the means available to suppress them.

This paper describes the optimal design of an \gls{smf}-coupled optical terminal for a range of \gls{fso} link lengths and turbulence conditions. An embedded, fast-steering, \gls{tt} mirror corrects for any angular offsets in the beam, providing first-order atmospheric turbulence suppression and fine target tracking. The design enables terrestrial \gls{fso} links of up to 10 km in a low complexity and cost effective manner, and is intended for sub-0.2\,m diameter apertures. This work has informed the design of optical terminals used by our research group to demonstrate \gls{fso} communications and frequency transfer to stationary \cite{Karpathakis2023, Gozzard2022} and dynamic targets \cite{McSorley2024, Walsh2022}.

\section{Theoretical background}

\subsection{Atmospheric turbulence}

The design process begins by quantifying the turbulence of the free-space link. Local temperature and pressure variations in the atmosphere occur down to millisecond timescales. These variations create turbulent cells known as eddies with varying refractive indices \cite{Strohbehn1978}. As a laser beam propagates through the atmosphere, it interacts with these eddies and experiences wavefront perturbations. Changes in the gradient of the wavefront is called wavefront \gls{tt}. Wavefront perturbations are quantified through the Fried parameter,\,$r_0$, which describes the spatial coherence diameter of the beam after propagation over a given \gls{fso} link. The larger a beam is relative to $r_0$, the more perturbed its wavefront will be. The simplest approximation of a propagating beam is a plane-wave, which has a Fried parameter given by $r_0 = 2.1 \rho_{0,pl}$, where $\rho_{0,pl}$ is the plane-wave spatial coherence radius given by 

\begin{equation}\label{eq:rho_plane}
   \rho_{0,pl} = \left [ 1.46 \left (\frac{2 \pi}{\lambda} \right )^2 \int_0^L dz \,C_n^2(z)\right ]^{-3/5}\;.
\end{equation}

\noindent Here $z$ is line-of-sight position along the link, $L$ is the link length, $\lambda$ is the beam's wavelength, and $C_n^2(z)$ is the refractive index structure parameter. For simplicity, this paper assumes a uniform distribution, $C_n^2(z) = C_n^2$. This is a suitable simplification for horizontal links, but for significantly inclined links, $C_n^2(z)$ distributions such as the Hufnagel-Valley model \cite{beland1993propagation} should be used.

Note that Equation \ref{eq:rho_plane} is only applicable to approximately plane-wave beams under weak turbulence conditions. However, this paper considers Gaussian beams over a range of turbulence conditions. For general turbulence conditions, the Fried parameter of a Gaussian beam is given\,by

\begin{equation}\label{eq:fried}
    r_0 = 2.1\rho_{0,pl} \left( \frac{8}{3 \left( a_e + 0.62 \Lambda_e^{11/6} \right )}\right) ^ {3/5}\:,
\end{equation}

\noindent where $\Lambda_e$ is the effective fresnel ratio beam parameter, $a_e$ is given by

\begin{equation}
    a_e = 
    \left\{ 
    \begin{array}{cc}
      \frac{1-\Theta_e^{8/3}}{1-\Theta_e} & \Theta_e \geq 0 \\
      \frac{1+{\left|\Theta_e\right|}^{8/3}}{1-\Theta_e} & \Theta_e < 0 \\
    \end{array}
\right. \:,
\end{equation}

\noindent and $\Theta_e$ is the effective curvature beam parameter. $\Lambda_e$ and $\Theta_e$ are modifications of the standard beam parameters, $\Lambda$ and $\Theta$, as defined in \cite{Andrews2005} to account for how the turbulent atmosphere impacts beam propagation. The effective beam parameters are functions of $C_n^2$, $z$ and $w_0$, the waist size of the transmitted beam, and equations for them can be found in \cite{Andrews2005}.

Wavefront perturbations may be represented through the Zernike modes. Compensation of a beam's perturbed wavefront up to the \textit{j}th Zernike mode, $Z_j$, will increase the beam's spatial coherence length. This compensation can be quantified through the generalised Fried parameter,\,$r_{0,j}$ \cite{Cagigal2000}

\begin{equation}\label{eq:generalised_fried}
    r_{0,j} = \left ( \frac{3.44}{\mathrm{coef(\textit{j})}} \right )^{3/5} 0.286 j^{-0.362} r_0\;,
\end{equation}

\noindent where $\mathrm{coef}(j)$ is the coefficient of residual wavefront distortion after partial compensation given by \cite{Noll1976}. Of interest to this paper are the second term $Z_2$ and third term $Z_3$, which represent wavefront tip and tilt. Equation\,\ref{eq:generalised_fried} can be used to give the \gls{tt}-compensated Fried parameter, ${r_{0,3} = 1.347 r_0}$, which is key in determining the improvements offered by \gls{tt} compensation. 

\subsection{Fibre-optic coupling}

Ultimately, wavefront perturbations will cause a loss in optical power when coupling a beam into \gls{smf}. For an incoming beam, the coupling efficiency, $\eta_c$, is the time-averaged fraction of incident power that is coupled into optical fibre. It is calculated from the overlap integral of the beam's field with the fibre's propagation mode. For a perturbed beam being focused onto an optical fibre which has fully filled some aperture diameter, $D_{RX}$, the coupling efficiency is given by \cite{Dikmelik2005}

\begin{equation}\label{eq:coupling_eff}
    \eta_c = 8 a^2 \int_0^1 \int_0^1 \exp \left [- \left (a^2+ 1.1 \left ( \frac{D_{RX}}{r_0} \right )^2 \right )
    (x_1^2 + x_2^2) \right ] J_1 \left (2.2 \left ( \frac{D_{RX}}{r_0} \right ) ^2 x_1 x_2 \right ) x_1 x_2 dx_1 dx_2\;,
\end{equation}

\noindent where $J_1$ is the first order Bessel function of the first kind, and $x_1$ and $x_2$ are normalised distances from the centre of the aperture stop. The parameter \textit{a} is determined by the geometry of the optics

\begin{equation}\label{eq:aparam}
    a = \frac{D_{RX}}{2}\frac{\pi W_m}{\lambda \, f_{\mathrm{eff}}}\;,
\end{equation}

\noindent where $W_m$ is the mode field radius of the fibre, $D_{RX}$ is the diameter of the aperture stop, and $f_{\mathrm{eff}}$ is the effective focal length of the optical system \cite{Dikmelik2005}. 

\begin{figure}[ht!]
    \centering
    \includegraphics[]{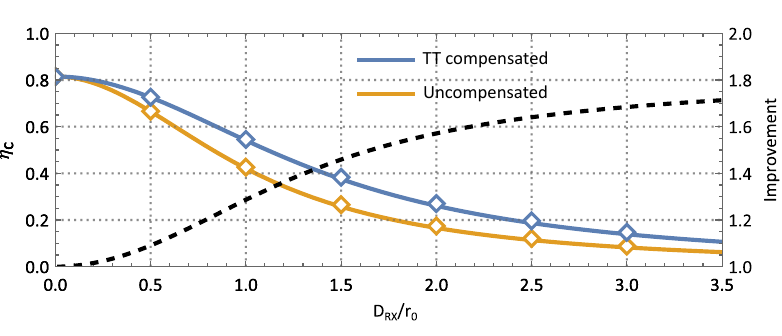}
    \caption{Coupling efficiency, $\eta_c$, versus turbulence strength, $D_{RX}/r_0$. The orange curve is for an uncompensated wavefront, and the blue is for a wavefront with \gls{tt} compensation. Both curves are evaluated for a = 1.12. The diamonds correspond to values of $a$ which maximise $\eta_c$ for a specific $D_{RX}/r_0$. The black dashed curve is the factor of improvement gained by \gls{tt} compensation.}
    \label{fig:couplingeff}
\end{figure}

In the absence of turbulence (where $D_{RX}/r_0 \rightarrow 0$), $\eta_c$ is maximised to 0.81 when $a = 1.12$. In the presence of turbulence, $\eta_c$ will decrease with increasing $D_{RX}/r_0$, shown as the blue curve in Figure\,\ref{fig:couplingeff}. For any value of $r_0$, setting $a = 1.12$ will yield a near maximum value of $\eta_c$, showcasing how useful \textit{a} is as a design parameter. The Fried parameter can be substituted with $r_{0,3}$ in Equation\,\ref{eq:coupling_eff} to find the \gls{tt}-compensated $\eta_c$, plotted in orange in Figure\,\ref{fig:couplingeff}. The fractional improvement in $\eta_c$ from \gls{tt} compensation increases with increasing turbulence strength, up to a value of $1.7$.

\subsection{Link power loss and fluctuations}

Only the fraction of power that makes it into the optical terminal can be coupled into fibre. For a given \gls{fso} link, this fraction is determined from the geometric loss associated with beam truncation at any aperture along the optical path. Geometric losses can be calculated by using link budget analysis. The net gain from transmitter to receiver, $G_{link}$, can be expressed through \cite{CarrascoCasado2020}

\begin{equation}\label{eq:G_link}
    G_{link} = \frac{G_{TX} G_{RX}}{L_{FS}}\;,
\end{equation}

\noindent where $G_{TX}$ is the transmitter gain, $G_{RX}$ is the receiver gain, and $L_{FS}$ is the free-space loss. For a transmitted Gaussian beam with a $1/e^2$ diameter of $D_{TX}$ that diverges by its diffraction limit, being received by a circular aperture of diameter $D_{RX}$ after propagating over a distance L, these three gain parameters are given by

\begin{equation}
    G_{TX} = 2 \left (\frac{\pi D_{TX}}{\lambda} \right )^2,\;\;
    G_{RX} = \left (\frac{\pi D_{RX}}{\lambda} \right ) ^2,\;\;  \text{and} \;\;
    L_{FS} = \left ( \frac{4 \pi L}{\lambda} \right ) ^2\;.
\end{equation}

Additionally to time-averaged power losses, the power received by the optical terminal will scintillate due to self-interference of the perturbed beam. These fluctuations are often quantified by the log-amplitude scintillation index

\begin{equation}
    \sigma_{ln,I}^2 = \ln \left(1+\frac{\sigma_I^2}{\mu_I^2} \right) \;,
\end{equation}

\noindent where $\sigma_I^2$ is the variance of the irradiance and $\mu_I$ is the mean irradiance. In general, $\sigma_{ln,I}^2$ will increase with increasing turbulence strength, reaching its maximum value when the beam has been perturbed to the point that it is fully incoherent at the receiver \cite{Clifford1974}. 

The ratio $D_{RX}/r_0$ has a strong effect on the scintillation at the receiver. If the aperture size spans multiple Fried parameters, the receiver will average the power fluctuations of each coherent patch, resulting in a lower scintillation index. This is termed aperture averaging \cite{Churnside1991}. We are further interested in how a varying aperture size affects the scintillation index of the power coupled into \gls{smf}. Given their analytical complexity \cite{Andrews2005}, we use numerical beam propagation simulations to obtain these scintillation indices.

\section{Numerical simulation}

We model the turbulent atmosphere with the AOtools Python package \cite{Townson2019} using the equivalent layer method \cite{Osborn2021}, where the volume of atmosphere over a \gls{fso} link is reduced to a series of thin phase screens, with Fresnel beam propagation between them. We use seven equivalent layers to model every turbulent link. This spacing yields a maximum error of 15\% compared to analytical solutions for scintillation of an unbounded plane-wave \cite{Andrews2005} under the strongest turbulence conditions and longest distances simulated.

The simulation parameters are chosen using the procedure detailed in \cite{Johnston2000}. The pixel scale is set small enough to ensure adequate sampling of the phase screens and of the beam scintillation. The grid size is set to minimise edge effects and allow for accurate Fresnel propagation. We note that after beam propagation, only a smaller central portion of the grid will contain valid results \cite{Johnston2000}.

\section{Terminal design}

\begin{figure}[ht!]
\centering
\includegraphics[]{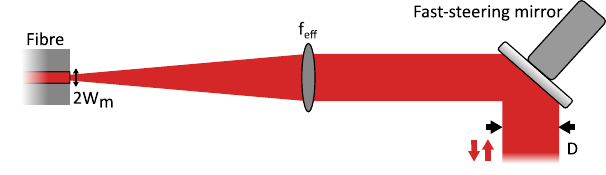}
\caption{Generalised optical terminal diagram. \textit{D} is the aperture stop, $f_{\mathrm{eff}}$ is the effective focal length, and $W_m$ is the mode field radius of the fibre.}
\label{fig:opticsystem}
\end{figure}

The generalised optical terminal is shown in Figure\,\ref{fig:opticsystem}. It receives a beam from free-space and focuses it down onto an optical fibre tip. A fast-steering mirror is used to keep the beam on the centre of the fibre tip, providing \gls{tt} compensation. The design is done in two stages. First is a maximisation of the total power throughput (i.e., the combination of coupling efficiency and geometric loss) and minimisation of scintillation assuming a perfectly centred beam. Second is an optimisation of the \gls{tt} compensation system to keep the coupling efficiency above some threshold value. Measurement of wavefront \gls{tt} to provide error signals for the steering control loop is not discussed. This is already well understood and detailed in publications such as \cite{DixMatthews2023a, Manning2015}.

\subsection{Power throughput optimisation}

\begin{figure}[ht!]
\centering
\includegraphics[]{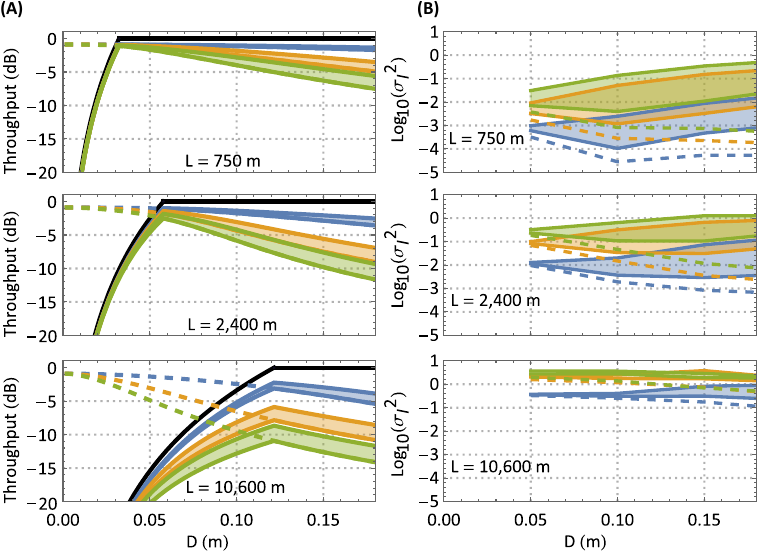}
\caption{(A) $G_{link}$ (black curve), uncompensated $\eta_c$ (dashed curves) and $\eta_{net}$ (solid curves) for three point-to-point links of length 750\,m, 2,400\,m, and 10,600\,m between a transmitter and receiver of equal aperture diameter. Three $C_n^2$ values of $1\e{-15}$ $m^{-2/3}$ (blue), $5\e{-15}$ $m^{-2/3}$ (orange) and $1\e{-14}$ $m^{-2/3}$(green) are used. The upper and lower throughput curves correspond to the \gls{tt}-compensated and uncompensated cases, and the filled region represents the range for imperfect \gls{tt} compensation. (B) Simulation results of the scintillation indices for free-space (dashed curve), and fibre (solid curves). The lower and upper fibre curves correspond to the \gls{tt}-compensated and uncompensated cases.}

\label{fig:throughput}
\end{figure}

The total power throughput, $\eta_{net}$, depends both on the coupling and geometric power losses of the \gls{fso} link through $\eta_{net} = G_{link} \eta_c$. For a given link, as $D_{TX}$ increases, the divergence of the transmitted beam decreases, and so geometric losses decrease. As $D_{RX}$ increases, the ratio $D_{RX}/r_0$ increases, implying more wavefront perturbations and hence a higher coupling loss. This paper considers the case where $D_{RX} = D_{TX} = D$, representative of bidirectional \gls{fso} terminals. In this case, there will be a common aperture diameter which will maximise $\eta_{net}$. This is plotted in Figure\,\ref{fig:throughput}A, showing $\eta_c$, $G_{link}$, and $\eta_{net}$ for a range of link lengths and $C_n^2$ values.

Three chosen values of $C_n^2$ representing the low, medium, and high turbulence conditions expected over each link are chosen based on the $C_n^2$ values our research group has observed over the specified link lengths in Perth, Western Australia. The total throughput in Figure \ref{fig:throughput} is given by a band, with the lower and upper bounds corresponding to the uncompensated and \gls{tt}-compensated cases. Each $\eta_{net}$ band has a peak value corresponding to an ideal choice for the aperture diameter, $D$. These peaks correspond to the minimum aperture size required for the beam to be minimally truncated at the receiver. The optimal $\eta_{net}$ ranges between $-1$\,dB to $-9$\,dB for the link lengths and turbulence strengths being analysed. In general, geometric losses are seen to have more impact on the $\eta_{net}$ than coupling losses do. As a result, the ideal $D$ is strongly dependent on the \gls{fso} link length. For lengths on the order of a kilometre, $D$ of less than 30 mm will incur significant throughput losses. The ideal $D$ varies from $30$ to $120$ mm over this range. 

With the aperture size set, Equation\,\ref{eq:aparam} can be used to set the terminal's effective focal length, resulting in a value between $152$ to $574$\,mm. The optimal aperture size and the resulting $\eta_{net}$ for each link and turbulence condition are summarised in Table \ref{tab:throughput_vals}.

Beyond a link length of 10 km, the optimal aperture sizes of >0.12\,m become inappropriate for a `small aperture' optical terminal. On telescopes this large, implementing only \gls{tt} compensation has diminishing improvements on the coupling efficiency, since wavefront perturbations are distributed into higher-order Zernike modes \cite{Chen2015}. As such, large aperture terminals generally require higher order compensation systems to improve coupling efficiency. This analysis does not consider higher order compensation, focusing on a lower cost and more simple design. 

\begin{table}[ht!]
\small
    \centering
    \caption{Optimal aperture diameter, D, corresponding required focal length, $f_{\mathrm{eff}}$, and resulting throughput $\eta_{net}$ for a given link length and turbulence condition. The range of $r_0$ values from Equation \ref{eq:generalised_fried} for the transmit beam sizes considered are also included.}
    \begin{tabular}{c|c|c||c|c|c}
        \hline
         L [m] & $C_n^2$ [$\mathrm{m^{-2/3}}$] & $r_0$ [mm]  & D [mm] & $f_{\mathrm{eff}}$ [mm] &$\eta_{net}$ [dB] \\ \hline
         \multirow{3}{*}{750}&$1\e{-15}$&$410-371$ &\multirow{3}{*}{32.4}&\multirow{3}{*}{152} &-0.899 \\
         &$\mathrm{5\e{-15}}$ &$158-141$ & & &-0.948 \\ 
         &$\mathrm{1\e{-14}}$ &$107-93.5$ & & &-1.02 \\ \hline
         \multirow{3}{*}{2,400}&$1\e{-15}$ &$280-187$ &\multirow{3}{*}{57.9} &\multirow{3}{*}{272} &-0.994 \\
         &$\mathrm{5\e{-15}}$ &$117-74.8$ & & &-1.43 \\
         &$\mathrm{1\e{-14}}$ &$8.42-5.27$ & & &-1.85 \\ \hline
         \multirow{3}{*}{10,600}&$1\e{-15}$ &$144-125$ &\multirow{3}{*}{122} &\multirow{3}{*}{574} &-2.24 \\
         &$\mathrm{5\e{-15}}$ &$5.9$8 & & &-5.90 \\
         &$\mathrm{1\e{-14}}$ &$3.99-4.00$ & & &-8.69  \\ \hline
    \end{tabular}
    \label{tab:throughput_vals}
\end{table}

Each link length and $C_n^2$ value is numerically simulated to obtain the free-space and fibre scintillation indices which are plotted in Figure \ref{fig:throughput}B. The free-space scintillation index is shown by a dashed line. As in the throughput plots, a band gives the upper and lower limits on the fibre scintillation index without and with \gls{tt} compensation, respectively. The uncompensated fibre $\sigma_{ln,I}^2$ is calculated from the fibre-coupled power, which is found by convolving the aperture-plane free-space power with the back-propagated fibre mode \cite{Chen2015}. The \gls{tt}-compensated fibre $\sigma_{ln,I}^2$ is found by centering the Fourier transform of the incident beam prior to convolution. This is equivalent to suppressing the G tilt of the beam, representative of how \gls{tt} compensation is performed when using a camera or quadrant photodetector \cite{Tyler1994}.

Assuming the free-space detector can see the entire aperture of the receiver, \gls{tt} compensation has no effect on the free-space $\sigma_{ln,I}^2$. Observing Figure \ref{fig:throughput}B, as the turbulence strength increases, all three scintillation indices increase, and when $D$ increases, aperture averaging effects predictably reduce the free-space scintillation. However, the scintillation in fibre is not as straightforward, since it depends on the scintillation of the incident beam and also on its wavefront perturbations. So, the net trend of fibre scintillation with aperture diameter depends on the balance of these two effects. We see that the scintillation index increases with increasing $D$ for lower turbulence strengths and this trend becomes weaker eventually changes sign as the turbulence strength increases. Compared to shorter links, the fibre and free-space scintillation indices of the 10,600\,m link do not change as much when $C_n^2$ increases. This is attributed to being close to/within the scintillation saturation regime \cite{Clifford1974}.

Looking at the improvements offered by \gls{tt} compensation, we see in Figure \ref{fig:throughput}B that \gls{tt} compensation has the largest improvement on fibre scintillation for lower turbulence strengths. This is because of how Zernike modes are distributed over different turbulence strengths. In weak turbulence, \gls{tt} is the dominant mode \cite{Chen2015} and so \gls{tt} compensation will remove a majority of perturbations from the beam, resulting in large improvements in the scintillation index. In stronger turbulence conditions, wavefront perturbations excite higher order Zernike modes, so only compensating for \gls{tt} has less of an impact. Additionally, as the free-space detector is assumed to be insensitive to \gls{tt}, weak turbulence conditions yield the biggest difference between the fibre and free-space\,$\sigma_{ln,I}^2$.

\begin{figure}[ht!]
    \centering
    \includegraphics[]{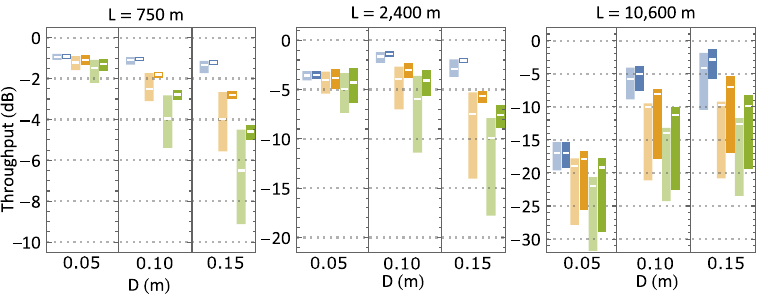}
    \caption{Upper and lower quartiles of $\eta_{net}$ for each link length and turbulence condition from Figure \ref{fig:throughput}A, for three different aperture diameters. Three $C_n^2$ values of $1\e{-15}$, $5\e{-15}$ and $1\e{-14}$ $m^{-2/3}$ are used, coloured blue, orange and green respectively. The light and solid bars correspond to the \gls{tt} uncompensated and compensated cases, respectively. The white dashes within each bar correspond to $\eta_{net}$ from Figure \ref{fig:throughput}.}
    \label{fig:powervariance}
\end{figure}

Note that the $D$ is already constrained to maximise the power throughput. However, using these $D$ values in Figure \ref{fig:throughput}B does not lead to a minimised fibre scintillation index. Minimising the fibre scintillation index requires $D$ in the range of $0.10-0.15$\,m, which would result in a lower throughput. Improved scintillation may be favoured over a reduction in throughput, though the overall power distribution in fibre is ultimately the most important metric when considering the operation of fibre-based \gls{fso} components such as detectors or amplifiers. The throughput distributions are presented in Figure \ref{fig:powervariance} with the inclusion of the upper and lower quartile of the total throughput from each simulation. The effect of \gls{tt} compensation is better highlighted here, showing significant reduction of throughput distribution's interquartile range for all turbulence strengths over the 750 and 2,400 m links. As discussed previously, the 10,600\,m link causes predominantly non-\gls{tt} wavefront perturbations, and so \gls{tt} compensation does little to the throughput distribution, except at sufficiently low $C_n^2$ values.

Most importantly, Figure \ref{fig:powervariance} shows that the \gls{tt}-compensated interquartile ranges are not strongly dependent on $D$. This confirms that choosing $D$ to optimise $\eta_{net}$ does not come at the cost of a sub-optimal fibre power distribution. Figure \ref{fig:powervariance} also reinforces that proper operation of the \gls{tt} compensation system is critical in achieving optimal fibre power distributions.

\subsection{Tip-tilt bandwidth optimisation}

With all the passive optical elements from Figure \ref{fig:opticsystem} now chosen, the design of the \gls{tt} compensation system may be examined. The system must suppress the wavefront \gls{tt} so the scintillation and total throughput can match the \gls{tt}-compensated cases. The two parameters of interest are the mirror's maximum angular range and the frequency response of the overall steering control loop. To constrain these parameters, the effect of beam misalignment on fibre coupling efficiency should be isolated. For the configuration shown in Figure\,\ref{fig:opticsystem}, any angular offset in the incoming beam of $\theta$ will result in a translation of the focused beam along the fibre tip equal to $f_{\mathrm{eff}}\theta$. By treating the incoming beam and propagation mode of the fibre as Gaussian, the resulting \gls{tt}-only coupling efficiency is given by \cite{Marcuse1977}

\begin{equation}\label{eq:etactilt}
    \eta_{c,TT}(\theta) = \left (\frac{2 \omega_1^2 \omega_2^2}{\omega_1^2+\omega_2^2} \right )^2 \exp \left [-\frac{2 \left (f\theta \right)^2}{\omega_1^2 + \omega_2^2}  \right ]\;,
\end{equation}

\noindent where $\omega_1$ and $\omega_2$ are the $1/e^2$ spot radius and mode field radius of the focused beam and fibre tip respectively. 

Equation\,\ref{eq:etactilt} can be simplified noting the following: the spot size of the focused Gaussian beam is $\omega_1 = 2\lambda f/\pi D$ from ray analysis; and the effective focal length $f_{\mathrm{eff}}$ is constrained via the \textit{a} parameter from Equation\,\ref{eq:aparam}. With this in mind, the isolated effect of \gls{tt} on coupling efficiency can be expressed through the normalised \gls{tt}-only coupling efficiency, $\Tilde{\eta}_{c,TT}(\theta) = \eta_{c,TT}(\theta)/\eta_{c,TT}(0)$

\begin{equation}\label{eq:etactiltnorm}
    \Tilde{\eta}_{c,TT}(\theta) = \exp \left [ -0.22 \left ( \frac{\pi D \theta}{\lambda} \right )^2  \right ]\;.
\end{equation}

We now obtain the expected atmospheric \gls{tt} variation, referred to as jitter, to use with Equation\,\ref{eq:etactiltnorm} to specify the \gls{tt} compensation system's required range and bandwidth. Jitter is determined by the \gls{tt} component of the overall atmospheric noise \gls{psd}, $S_{TT}(f)$. This is approximated by \cite{Tyler1994} to be

\begin{equation}\label{eq:S_tilt}
    S_{TT}(f) = 0.31 D^{-2} f^{-8/3} \int_L dz \, {C_n}^2(z) V(z)^{5/3} F_G\left(f D/V(z)\right)\;,
\end{equation}

\noindent where $V(z)$ is the perpendicular wind speed and

\begin{equation}
    F_G(y) = \int_0^1 dx \, \frac{x^{5/3}}{\sqrt{1-x^2}} {J_1}^2(\pi y/x)\;.
\end{equation}

Equation\,\ref{eq:S_tilt} can be integrated to find the jitter variance, $\sigma_{jitter}^2$ \cite{Tyler1994}. The designed maximum optical range of the fast-steering mirror is then set based on the square root of this value. We choose a range of $3\sigma_{jitter}$, corresponding to $5$ to $75$ $\mu$rad for the range of turbulence strengths and optimal aperture sizes being analysed. 

The bandwidth of the steering control loop dictates its frequency response, $H(f)$. In the simplest approximation, $H(f)$ is given by

\begin{equation}\label{hsteering}
    H(f) = \frac{1}{1+if/f_{3dB}}\;,
\end{equation}

\noindent where $f_{3dB}$ is the bandwidth, imposed by processing delays and actuator dynamics. The residual \gls{tt} spectrum is then given by  $\left| 1 - H(f) \right| ^2 S_{TT}(f)$, which can be integrated to give the residual jitter variance, $\sigma_{jitt,res}^2$.

\begin{figure}[ht!]
    \centering
    \includegraphics[]{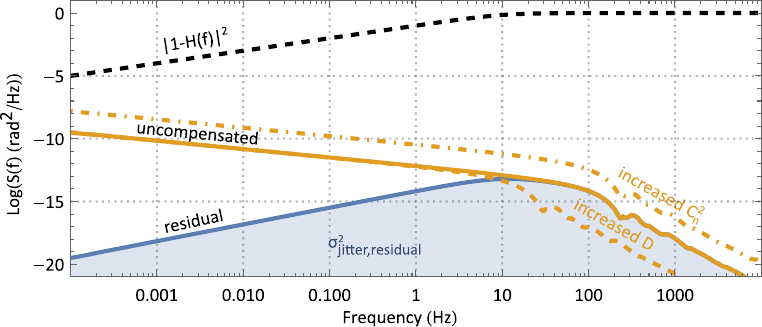}
    \caption{Uncompensated (orange) and residual (blue) atmospheric \gls{tt} spectra for a suppression bandwidth of 10 Hz. The orange dot-dashed and dashed curves show the effect of increasing the aperture size and turbulence strength respectively. The fast-steering mirror's residual transfer function is shown as a black, dotted curve. The integrated residual jitter spectra gives the residual jitter variance, $\sigma_{jitt,res}^2$.}
    \label{fig:jitterspectra}
\end{figure}

Figure\,\ref{fig:jitterspectra} shows an example of uncompensated and residual \gls{tt} spectra for a control loop bandwidth of 10 Hz. The fast-steering mirror is characterised by diminishing suppression as the bandwidth of the control loop is approached. The uncompensated spectrum shifts from an $f^{-2/3}$ to $f^{-11/3}$ frequency dependence past the frequency $f_{knee} = V/D$, due to the effects of aperture averaging \cite{Tyler1994}. Hence, to suppress most of the jitter in a \gls{fso} link, the bandwidth of the control loop should at least be equal to $f_{knee}$. Figure\,\ref{fig:jitterspectra} also shows the effect of aperture size and turbulence strength on the \gls{tt} spectrum through the dashed and dot-dashed orange curves respectively. An increase in aperture size will decrease $f_{knee}$, reducing the uncompensated jitter, whereas an increase in turbulence strength will uniformly increase the magnitude of the entire \gls{tt} spectrum.

To quantify the actual \gls{tt} suppression required, the residual jitter should keep the normalised \gls{tt}-only coupling efficiency above 90\%, a choice inline with that presented in \cite{Tyler1994}. This constraint will bring the total throughput and scintillation close to the \gls{tt}-compensated case from Figure\,\ref{fig:throughput}. Using Equation\,\ref{eq:etactiltnorm}, this requirement constrains the bandwidth like so:

\begin{equation}
      \left ( \int df \, \left| 1 - H(f) \right| ^2 S_{TT}(f) \right )^{1/2} < 0.22 \frac{\lambda}{D}\;.
\end{equation}

\begin{figure}[ht!]
    \centering
    \includegraphics[]{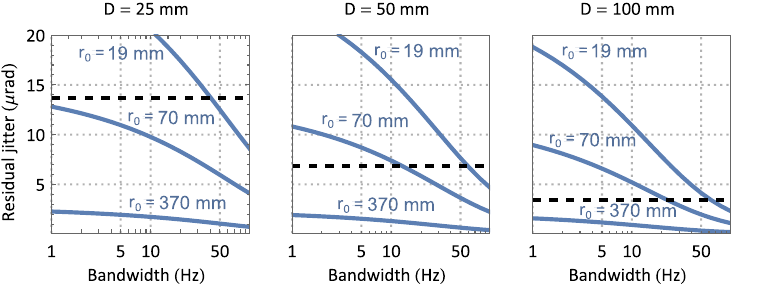}
    \caption{Residual jitter values as a function of \gls{tt} suppression bandwidth for different aperture diameters. Three blue curves for difference turbulence conditions (Fried parameters) are shown. The dashed black curves correspond to the maximum residual jitter permitted for different $D$ so that $\Tilde{\eta}_{c,TT}$ remains above 90\%.}
    \label{fig:residualjitter}
\end{figure}

Figure\,\ref{fig:residualjitter} shows how $\sigma_{jitt,res}$ varies with the control loop bandwidth. It highlights the minimum bandwidth needed to meet the specified value of $\Tilde{\eta}_{c,TT}(\sigma_{jitt,res})$ for a range of practical aperture sizes and fried parameters. The required bandwidth ranges from $1$ to $70$ Hz across all configurations, and particularly weak turbulence strengths do not require \gls{tt} compensation at all. In general, a larger aperture size will lead to less residual jitter but also make $\Tilde{\eta}_{c,TT}(\theta)$ more sensitive to jitter. The net result is that the required bandwidth increases with increasing aperture size. This behaviour opposes what Figure\,\ref{fig:jitterspectra} suggests and highlights the importance of using the normalised \gls{tt}-only coupling efficiency to inform the steering control loop bandwidth. 

Table \ref{tab:tt_bandwidth} shows the required \gls{tt} compensation bandwidth and range for each link when using the optimal aperture diameter, $D$, identified in Figure \ref{fig:throughput}A. Note that being passively below the $\sigma_{jitt,res}$ requirement does not suggest that \gls{tt} compensation is not required, but rather that a low ($\sim1$ Hz) bandwidth will suffice. In practice \gls{tt} compensation is always needed to correct for long-term path varying effects. As such, any sub-1 Hz bandwidths have been entered as <1 Hz in Table\, \ref{tab:tt_bandwidth}.

\begin{table}[ht!]
\small
    \centering
    \caption{Required \gls{tt} bandwidth, $f_{3dB}$, and range, $3\sigma_{jitter}$, for a given link length and turbulence condition based on the optimal aperture size, $D$, from Table \ref{tab:throughput_vals}.}
    \begin{tabular}{c|c||c|c|c}
        \hline
         L [m] & $C_n^2$ [$\mathrm{m^{-2/3}}$] & D [mm] & $f_{3dB}$ [Hz] & $3\sigma_{jitter}$ [$\mu rad$] \\  \hline
         \multirow{3}{*}{750}&$1\e{-15}$ & \multirow{3}{*}{32.4} & <1 & 7.68\\
         &$5\e{-15}$ & & <1 & 17.2\\ 
         &$1\e{-14}$ & & <1 & 24.2 \\ \hline
         \multirow{3}{*}{2,400}&$1\e{-15}$ & \multirow{3}{*}{57.9} & <1 & 12.5\\
         &$5\e{-15}$ & & 4.74 & 17.8\\
         &$1\e{-14}$ & & 15.8 & 39.4\\ \hline
         \multirow{3}{*}{10,600}&$1\e{-15}$ & \multirow{3}{*}{122} & 11.9 & 23.0\\
         &$5\e{-15}$ & & 42.3 & 51.5\\
         &$1\e{-14}$ & & 63.1 & 72.9\\ \hline
    \end{tabular}
    \label{tab:tt_bandwidth}
\end{table}

For the case of dynamic \gls{fso} links, residual jitter arising from imperfect target tracking via a telescope mount or similar is typically much larger than that due to atmospheric turbulence \cite{Walsh2022, McSorley2024}. As a result, the modest improvement in coupling efficiency after \gls{tt} compensation shown in Figure\,\ref{fig:couplingeff} will be much more substantial for dynamic links, but also require a greater fast-steering mirror range. The required bandwidth should not change significantly as tracking errors are typically lower-frequency than atmospheric \gls{tt}. The \gls{tt} spectrum of a dynamic link can be obtained using a quadrant photodetector or camera \cite{Manning2015, DixMatthews2023a}. Once this is done, the preceding analysis can be used to constrain the \gls{tt} compensation bandwidth, yielding a coupling efficiency very close to the \gls{tt}-compensated curves in Figure\,\ref{fig:throughput}.

It should be clarified that the value of $\Tilde{\eta}_{c,TT}(\sigma_{jitter})$ is not an indication of the achievable increase in coupling efficiency if all \gls{tt} is compensated. While $\eta_c$ can be decomposed into contributions from Zernike modes via Equations \ref{eq:generalised_fried} and \ref{eq:coupling_eff}, the relationship between the residual jitter and the corresponding $Z_2$ and $Z_3$ coefficients is unclear. As such, it is difficult to relate an arbitrary amount of jitter to $\eta_c$ when other wavefront perturbations are present. This means the uncompensated $\eta_c$ for a dynamic \gls{fso} link is not producible with this analysis. However, as the \gls{tt}-compensated curves from Figure\,\ref{fig:throughput}A remains valid, following this analysis for a dynamic \gls{fso} link will still properly constrain the \gls{tt} compensation system to achieve the optimal throughput.

\section{Conclusion}

In this paper, we have outlined the process for designing a low-complexity optical terminal for fixed or dynamic terrestrial \gls{fso} links of lengths on the order of kilometres. The design balances geometric and fibre coupling losses by constraining the terminal's aperture size to achieve maximal power throughput over the link and minimal scintillation. The terminal features \gls{tt} compensation through a fast-steering mirror with a constrained range and bandwidth which keeps the power throughput and scintillation optimised in the presence of atmospheric turbulence and tracking errors. Optimised terminal parameters -- aperture diameter, effective focal length, total power throughput and \gls{tt} bandwidth and range -- are presented in Tables \ref{tab:throughput_vals} and \ref{tab:tt_bandwidth} for a variety of link lengths and turbulence strengths. For dynamic links, it is necessary to know the residual \gls{tt} due to imperfect target tracking, at which point the analyses presented are immediately applicable.

\begin{backmatter}
\bmsection{Funding} SmartSat Cooperative Research Centre (P1-18)

\bmsection{Acknowledgments} This work has been supported by the SmartSat CRC, whose activities are funded by the Australian Government’s CRC Program. A.F. is supported by an Australian Government Research Training Program Scholarship and a top-up scholarship funded by the Government of Western Australia. The authors thank Nicolas Maron and Skevos Karpathakis for manuscript feedback. 

\bmsection{Disclosures} The authors declare no conflicts of interest.

\end{backmatter}


\bibliography{TerminalPaper}

\end{document}